\documentclass{anabs}
\usepackage{psfig}
\usepackage{times}
\def\cgs{{erg cm$^{-2}$ s$^{-1}$}}
\def\ltsima{$\; \buildrel < \over \sim \;$}
\def\simlt{\lower.5ex\hbox{\ltsima}}
\def\gtsima{$\; \buildrel > \over \sim \;$}
\def\simgt{\lower.5ex\hbox{\gtsima}}
\Volume{1-2}              
\Year{2003}              
\Month{02}               
\Pagespan{000}{000}      

\begin{document}
\lhead[\thepage]{A.N. Author: Title}
\rhead[Astron. Nachr./AN~{\bf 324} (2003) 1/2]{\thepage}
\headnote{Astron. Nachr./AN {\bf 324} (2003) 1/2, 000--000}

\title{An XMM-Newton Study of Hard X-ray Sources}

\author{E. Piconcelli\inst{1,2}, L. Bassani\inst{1}, M. Cappi\inst{1}, F. Fiore\inst{3}, G. Di Cocco\inst{1}, J.B. Stephen\inst{1}}
\institute{IASF/CNR, Via Gobetti 101, I--40129 Bologna, Italy
\and 
Dipartimento di  Astronomia, Universit\'a di Bologna, via Ranzani 1, I--40127 Bologna, Italy
\and
Osservatorio di Roma, via Frascati 33, I--00044 Monteporzio, Italy}

\correspondence{piconcelli@bo.iasf.cnr.it}

\maketitle

\section{Scientific goal}
The advent of new X-ray telescopes on board XMM-Newton and Chandra satellites are improving our knowledge on the high energy astrophysics. Because of their unprecedented sensitivity and angular resolution they allow to perform accurate imaging and spectroscopic studies as ever made before (see Mushotzky 2002 for a review).
Concerning the Cosmic X-ray background (CXB) extreme deep X-ray surveys carried out with XMM-Newton and Chandra have resolved into point sources for more than 90\% and 80\% of its soft (0.5-2 keV) and hard (2-10 keV) X-ray flux respectively (Hasinger 2002). 
However such pencil-beam surveys are biased towards very faint sources at fluxes $F_{2-10}\sim$ 10$^{-16}$--10$^{-15}$ \cgs which provide too few photons to perform a meaningful spectral analysis.
In order to explore the spectral properties of these sources we have started a wide-angle spectroscopic survey at an intermediate 2-10 keV flux level (i.e. $\sim$ 10$^{-14}$ \cgs).
Our selection criteria enables us to perform a detailed spectral analysis for all the objects in the sample as well as to obtain good constraints on the predictions of synthesis models of the CXB (see Piconcelli et al. 2002, hereafter Paper I).
To date, our sample consists of 90 objects taken from 12 XMM-Newton observations: results for the first 7 {\it EPIC} fields, which includes 41 X-ray sources, are discussed in Paper I. We present here the preliminary results concerning the whole sample including five new deeper XMM-Newton measurements. A complete and detailed discussion will be reported in a forthcoming paper (Piconcelli et al. 2002b, in prep.).   
\section{Preliminary results}
We have 37 X-ray sources optically identified to date, most of which are broad line AGNs: except for a galaxy at z=0.0058 (i.e. NGC4291), the values of z span from $\sim$0.1 to $\sim$2. The 2-10 keV fluxes of the 90 sources range from 1 to 82 $\times$ 10$^{-14}$ \cgs and the luminosities span from $\sim$10$^{40}$ to $\sim$10$^{45}$ erg s$^{-1}$. We divided our sample in two subsamples: the BRIGHT sample with 42 X-ray sources with  $F_{2-10}\geq$ 5  $\times$ 10$^{-14}$ \cgs and the FAINT sample with 22 sources with 2 $<F_{2-10}<$ 5  $\times$ 10$^{-14}$ \cgs.  
In Fig.~1({\it left}) we compare the average spectral indices obtained for the BRIGHT ($\langle\Gamma\rangle$=1.54$\pm$0.03) and FAINT ($\langle\Gamma\rangle$=1.56$\pm$0.05) samples using a power-law plus Galactic absorption model with 2 previous ASCA and a recent Chandra stacked spectra analysis results. Fig.1({\it right})  shows the expectations from the synthesis model of the CXB (Comastri et al. 2001) and the fractions of absorbed objects (i.e. with N$_H>$ 10$^{22}$ cm$^{-2}$) obtained from the main hard X-ray surveys performed so far together with our results from BRIGHT and FAINT subsamples.

We conclude that our results indicate that the fraction of observed obscured objects ($\sim$30\%) is below the predictions (\simgt50\%) of popular models for the origin of CXB (see also discussion in Paper I). 

\begin{figure}[ht]
\vspace{-0.3cm}
\centerline{
\psfig{file=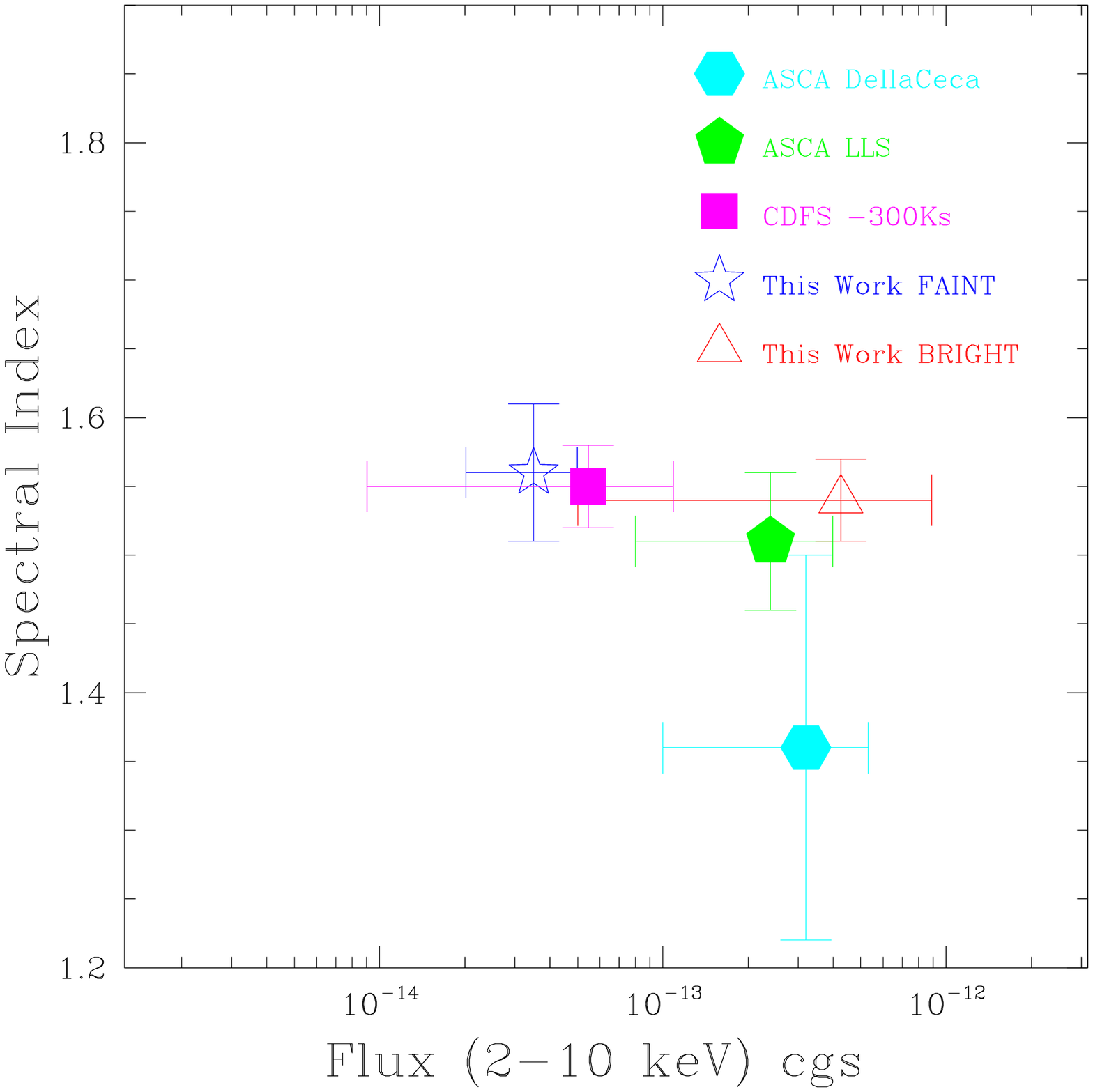,height=4.6cm,width=4.4cm,angle=0} 
\psfig{file=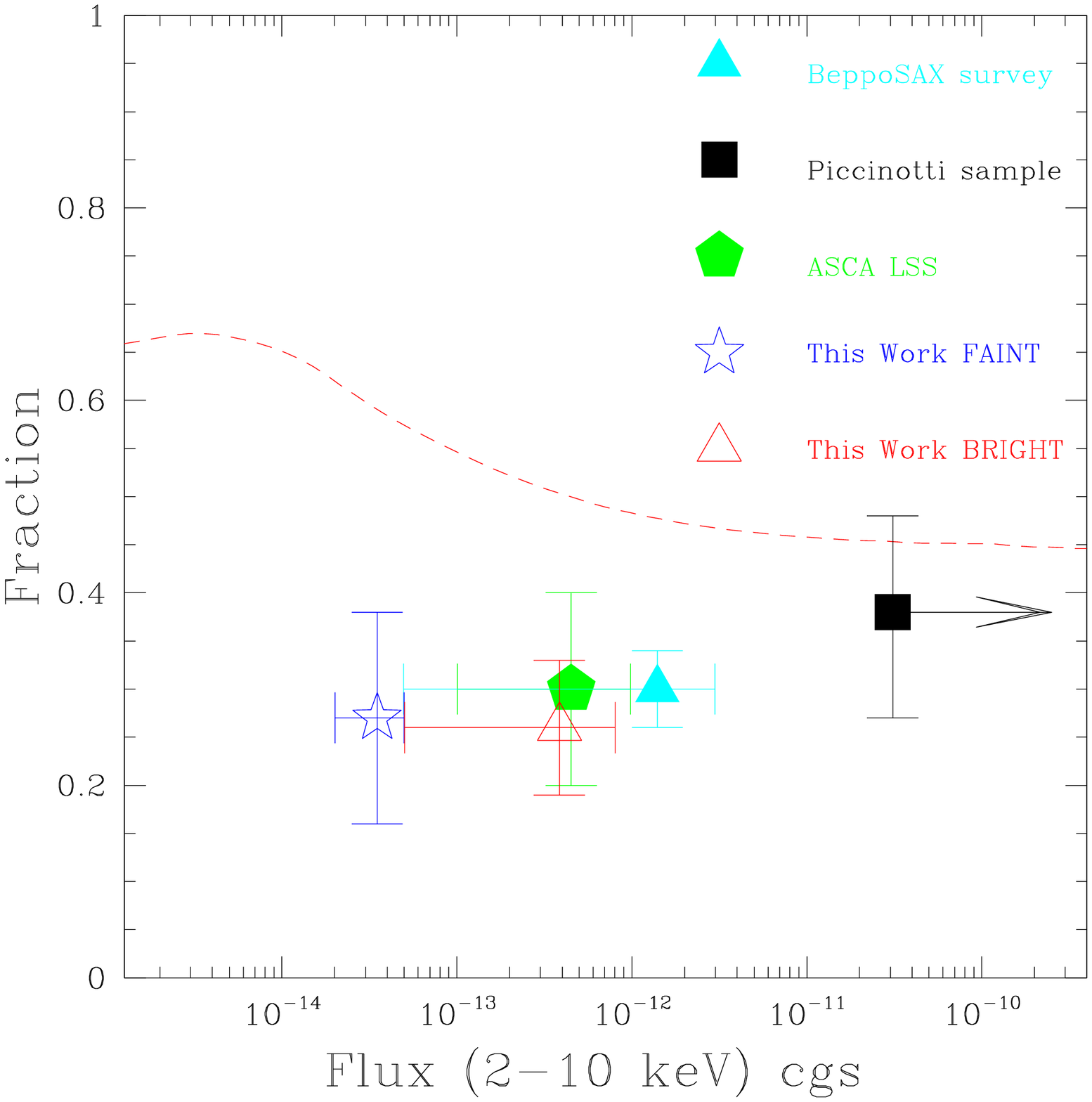,height=4.6cm,width=4.4cm,angle=0}
}
\caption{Comparison of our results with other hard X-ray surveys. Left: $\langle\Gamma\rangle$ versus 2-10 keV flux. Right: fraction of absorbed sources compared to predictions of synthesis models ({\it dashed line}, from Comastri et al. 2001).}
\label{hard_survey}
\end{figure}


\vspace{-0.9cm}
\begin{thebibliography}{}
\bibitem{} Comastri, A., et al.:2001, MNRAS 327, 871
\bibitem{} Hasinger, G.:2002 (astro-ph/0202430)
\bibitem{} Mushotzky, R.F.:2002 (astro-ph/0203310)
\bibitem{} Piconcelli, E., et al.: 2002, A\&A in press, (astro-ph/0208183)
\end{thebibliography}
\end{document}